# VLC Systems with CGHs

Safwan Hafeedh Younus[1], Ahmed Taha Hussein[1], Mohammed T. Alresheedi[2] and Jaafar M. H. Elmirghani[1]

[1]School of Electronic and Electrical Engineering, University of Leeds, Leeds, LS2 9JT, UK
[2]Department of Electrical Engineering, King Saud University, Riyadh, Saudi Arabia
E-mail: {elshy@leeds.ac.uk, asdftaha@yahoo.com, malresheedi@ksu.edu.sa, elshy@leeds.ac.uk, j.m.h.elmirghani@leeds.ac.uk}

**Abstract:** The achievable data rate in indoor wireless systems that employ visible light communication (VLC) can be limited by multipath propagation. Here, we use computer generated holograms (CGHs) in VLC system design to improve the achievable system data rate. The CGHs are utilized to produce a fixed broad beam from the light source, selecting the light source that offers the best performance. The CGHs direct this beam to a specific zone on the room's communication floor where the receiver is located. This reduces the effect of diffuse reflections. Consequently, decreasing the intersymbol interference (ISI) and enabling the VLC indoor channel to support higher data rates. We consider two settings to examine our propose VLC system and consider lighting constraints. We evaluate the performance in idealistic and realistic room setting in a diffuse environment with up to second order reflections and also under mobility. The results show that using the CGHs enhances the 3dB bandwidth of the VLC channel and improves the received optical power.

1. Introduction

Visible light communication (VLC) systems have experienced an exponential growth in interest because of the spectrum crunch and an increase in demand for high data rates. Currently, radio frequencies (RFs) are utilized to transfer the data. Unfortunately, the growth in demand for more frequencies, substantial bandwidth, higher data rates and better quality of service has faced limitations in terms of the availability of the RF spectrum [1]. Recently, a study by Cisco showed that data traffic is expected to rise tenfold by 2030 [2]. Compared with RF systems, VLC systems have license free and abundant bandwidth. Moreover, VLC systems are secure and have front-end devices with low-cost. One of the main challenges that face VLC systems, is situations where the line-of-sight (LOS) component is absent, which leads to severe impact on the VLC system performance. Furthermore, because of multi-path propagation in indoor VLC systems, inter-symbol interference (ISI) can be experienced, which limits the VLC system's achievable data rate. The former is a limitation in VLC systems, and is attributed to the channel properties [3], [4], where the absence of the line-of-sight (LOS) link increases the path loss [5], [6] and consequently degrades the performance of VLC systems.

Several studies have proposed approaches for reducing the impact of ISI in VLC systems. In [7], cooperative transmission was used to synchronize the time of arrival (TOA) of multiple LEDs, which leads to a reduction in ISI in VLC systems. LED spotlights were used with narrow half power angle to reduce the delay spread in an indoor VLC system and to increase the data rate [8]. A holographic light shape diffuser (LSD) was used to reduce the ISI and to obtain a uniform power distribution over the entire room [9]. An angle diversity receiver (ADR) was studied in indoor VLC systems to reduce the delay spread and consequently, reduce the impact of ISI and boost the system's performance [10], [11]. An imaging receiver with a delay adaptation technique (DAT) was also utilized to reduce the delay spread of an indoor VLC system. This enabled the system to work at high data rates with good performance [12]. Another approach used to decrease ISI is the use of CGH in indoor VLC systems [13], [14], [15], [16]. Two VLC systems were proposed based on the CGHs in [16]: a static CGHs-VLC system and an adaptive CGHs-VLC system where the CGHs was utilized to generate 100 fixed narrow beams in the static CGHs-VLC system and 8 steered beams in the adaptive CGHs-VLC system. However, these VLC systems were complex as many beams are generated by the CGHs and the exact position of the receiver has to be identified for proper operation.

Here, we propose a fixed CGHs concept used to generate a single fixed wide beam, which reduces the complexity of the VLC system, decrease the impact of ISI and boost the received optical power in indoor VLC systems. The CGH is utilized to orient 30% of the total output optical power from the best light source and focus this power on a small surface of 2 m × 2 m. Each light source (transmitter) is followed by a CGH. But, the data is sent through one chosen light source, which is the best light source that offers high optical power and low ISI. We designed our proposed communication system while considering acceptable lighting level in the environments and the impact of multipath propagation. Additionally, the proposed systems are examined in two different rooms under user mobility and using on-off-keying (OOK) modulation.

The paper is structured as follows. The VLC system model is given in Section 2. The design of the CGH is developed in Section 3. The evaluation of the proposed VLC systems and the results are outlined in Section 4. Section 5 gives the conclusions.

## 2. VLC System Model

To characterize the indoor VLC channel, two room scenarios are considered an idealistic room and a realistic room, which are denoted as room A and room B, respectively, as shown in Figs. 1 (a) and (b). Both rooms have areas of 8 m × 4 m (length × width) and heights of 3 m. Experimental measurements have shown that light rays reflect from plaster walls roughly in a form of the Lambertian beam [17]. Thus, reflective surfaces (walls, ceiling and floor) of room A and room B were assumed to reflect the optical signal in the form of a Lambertian pattern with 0.8 reflection coefficient for the ceiling and walls and 0.3 reflection coefficient for the floor [18]. In this paper, we took into account reflections up to the second order. This is because higher order reflections (third order reflections and beyond) have a small impact on the indoor VLC system performance [19]. Here, to model reflections, a ray-tracing algorithm is used in which the walls, ceiling and floor of the rooms were split into small surface elements. Each surface element is square-shaped and has an area $d_A$ with a reflection coefficient $\rho$. In addition, these surface elements are considered as Lambertian reflectors with $n = 1$, where $n$ is the Lambertain emission order. Although, higher accuracy can be achieved when the size of the surface elements is reduced, this leads to exponential rise in the computation time. Therefore, to keep computations within a reasonable time, in the first order reflections, a reflective element size of 5 cm × 5 cm was selected while a size of 20 cm × 20 cm was selected for surface elements in the second order reflections [20]. Room B is considered as an office environment which contains furniture, chairs, a door, windows, bookshelves and mini cubicle partitions (see Fig. 1 (b)). The reflection coefficients of the windows and the door were set to zero. Moreover, two walls of room B ($y$ = 8 m and $x$ = 4 m) contains bookstands and filling cabinets (see Fig. 1 (b)) and have reflection coefficients of 0.4 [21]. In addition, the furniture inside the realistic room was given a reflection coefficient similar to that of the floor (0.3).

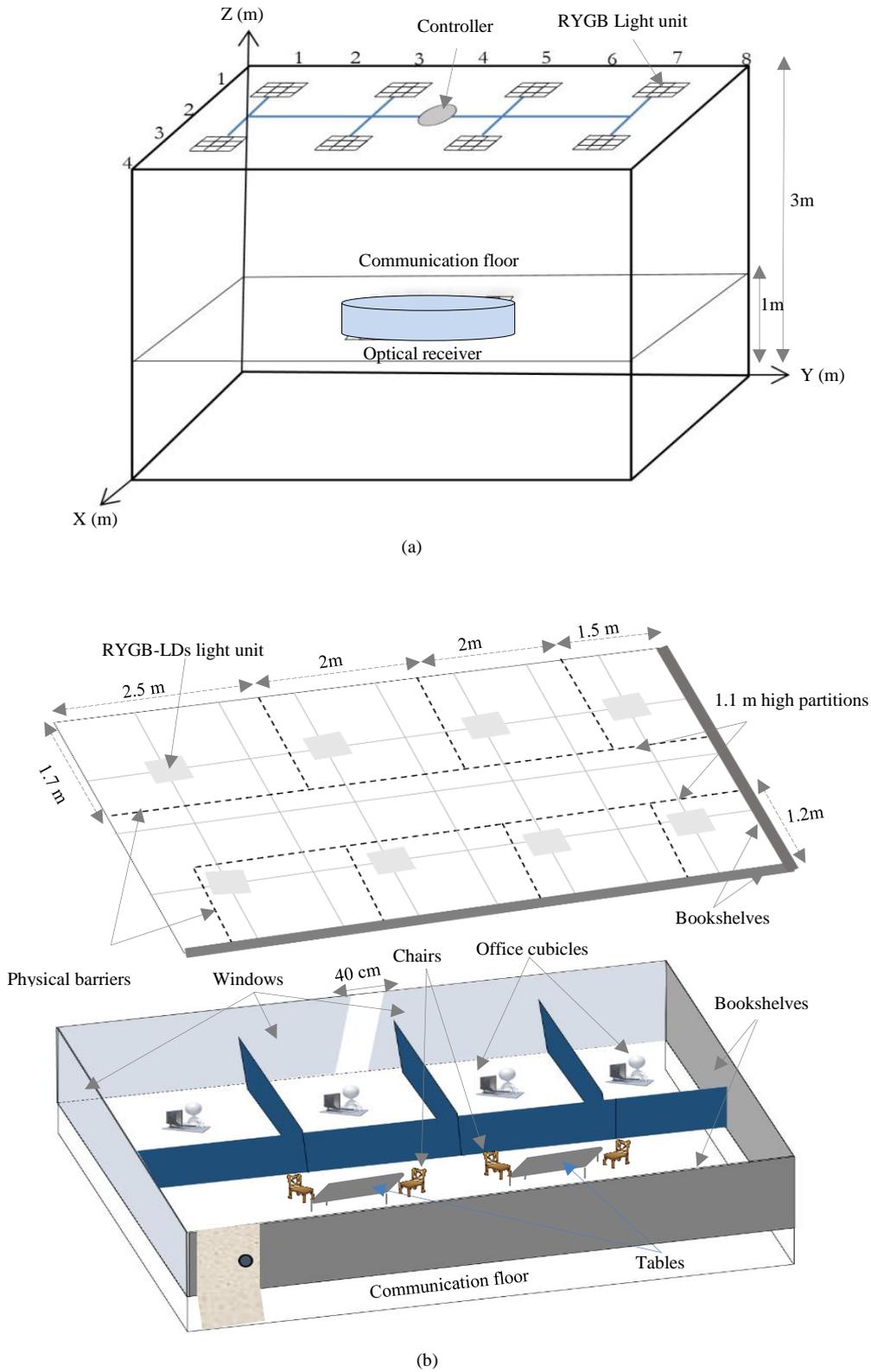

Fig 1: Room configurations (a) an idealistic environment (room A) and (b) a realistic environment (room B).

In this work, we considered laser diodes (LDs) as sources of lighting and as transmitters instead of light emitting diodes (LEDs). LDs offer many benefits over LEDs such as higher modulation speed, much brighter emissions compared to LEDs and higher output powers, which lead to high lumen output [22], [23]. In addition, LD illumination (four colours LD; RYGB)) can be designed to comply with human eye hazards regulations [24]. As such, we assumed the same parameters of the RYGB LDs in [24]. To guarantee a suitable level of illumination inside the proposed rooms following the ISO and European illumination framework [25], eight light sources are fitted on the room's ceiling (see Fig. 1 (a)). Each one of these luminaires contains nine RYGB LDs (3 × 3) with a 2 cm separation. It worth noting that the transmitted data modulated the four colours of the RYGB LDs and all RYGB LDs in the same light source carry the same data.

A wide field of view (FOV, FOV = 40º) detector with a photosensitive area of 0.5 mm$^2$ is proposed as a receiver front-end in this work. The wide-FOV receiver is the basic receiver configuration broadly studied, for example in [17], [26], [27]. The receiver height was 1 m, on the "communication floor" (see Fig. 1 (a)).

### 3. CGHs for Indoor VLC Systems

To attain an acceptable lighting level in the room, a number of spatially separated light sources are fitted in the room's ceiling. This produces ISI due mainly to the many strong LOS optical rays that arrive at receiver at different times. Therefore, in this work, the data is sent through only one light source. This light source is the one that provides the highest signal to noise ratio (SNR) at the optical receiver. But, this decreases the optical power that received by the optical receiver. Therefore, we used the fixed CGHs to direct a portion of the total optical power of the best light source and focusing it on a small surface area of the communication floor. Given the impact of diffuse reflections on the VLC system's performance, using the CGH reduces the impact of multipath propagation at higher data rates.

*3.1 Select the Best Light Source Algorithm.*

To reduce the effect of the LOS components from multiple light units that reach the optical receiver dispersed in time, one light source is utilized for data communication. A "select the best light source" (SBLS) algorithm is utilized following these steps:

1. Each light source is identified by an ID, and this ID is given by the controller.
2. The controller turns on the light sources one by one, which prevents interferences between the light sources.
3. The optical receiver obtains the SNR of each light source.
4. The optical receiver notifies the controller of the associated SNR of each light source by transmitting a feedback signal at low data rate. This feedback signal is assumed to be an infrared (IR) and we considered the design of IR uplink in [28].
5. The controller chooses the light source that offers the best performance (i.e., the light source that has the highest SNR) to transmit the data and deactivates data communications for the other light sources.

In some receiver locations, more than one light source can offer the same SNR, and this is related to the room's symmetry. The controller selects one light source and discards the other(s). It is worth observing that there is no need to find the optical receiver location while using the SBLS algorithm.

### 3.2 Design of the CGH for Indoor VLC Systems.

The CGHs have the ability to generate spots with any phase distribution and prescribed amplitude. The CGHs complex transmittance is used to compute the CGHs, which is given as [21]:

$$H(u,v) = A(u,v)e^{j\Phi(u,v)} \quad (1)$$

where $A(u,v)$, $\Phi(u,v)$ and $(u,v)$ denote the hologram amplitude distribution, the hologram phase distribution and the coordinates of the hologram in the frequency domain, respectively. The phase and the amplitude of the incoming wave-front can be modulated by the CGHs. Here, we used the CGHs to focus a specific amount of the best light source total optical power onto a small surface area of the communication floor. To compute the beam distribution, diffraction theory is utilized [29] to determine the phase of each pixel to get the wanted far-field pattern. While the hologram is in the frequency domain the observed far-field pattern is in the spatial domain. Hence, there are two domains, the frequency domain (the hologram domain) and the spatial domain (the far-field pattern domain), and a Fourier transform connects these two domains as [30]:

$$h(x,y) = \iint H(u,v)\, e^{-j2\pi(ux+vy)}\, dx\, dy \quad (2)$$

The hologram consists of an array ($M \times N$) of rectangular cells. The size of each cell is R × S and the value of the complex transmittance is $H_{kl}: -M/2\, k \leq M/2$ and $-N/2\, l \leq N/2$ [21]. The diffraction pattern of the hologram is written as [30]:

$$h(x,y) = RS\, sinc(R_x, S_y) \sum_{k=-\frac{M}{2}}^{\frac{M}{2}-1} \sum_{l=-\frac{N}{2}}^{\frac{N}{2}-1} H_{kl}\, e^{j2\pi(Rxk+Syl)} \quad (3)$$

A cost function (CF) is used to describe the difference between the actual output pattern and the desired far-field pattern. In the design of the CGHs, we used a simulated annealing algorithm for optimizing the CGHs. In this algorithm, the CGHs phase is gradually altered to get the wanted far-field pattern [31]. The pattern distribution in the far-field is $f(x,y) = |f(x,y)|\, e^{j\Phi(x,y)}$. The target of the CGHs design is to get the distribution of the CGHs $g(v,u)$ that generates a reconstruction $g(x,y)$ close to the desired distribution $f(x,y)$. The CF is the mean square error of the difference between the normalized wanted object energy $f''(x,y)$ and the scaled reconstruction energy of the $k^{th}$ iteration $g''(x,y)$ and it is written as [14]:

$$CF_k = \sqrt{\sum_{i=1}^{M} \sum_{j=1}^{N} (|f''(x,y)|^2 - |g''(x,y)|^2)^2} \quad (4)$$

### 3.3 Effect of the CGH on the Lighting Level.

Light engines in VLC produce very broad beams. In our case, beams with semi-angle of 70° were produced to meet the limitation requirements. These broad beams reduce the amount of power collected by the receiver, especially if only one light engine is used for communication as we propose here to reduce the ISI. Therefore, our CGHs direct part of the VLC light engine output power (30% in this work, optimized) towards the communication floor thus increasing the power collected by the optical receiver while maintaining illumination at levels acceptable by the standards.

Fixed CGHs were utilized in this work, where the CGHs produces a single wide fixed beam, (without steering to the receiver). Thus, the beam's size must be chosen to cover all the optical receiver possible locations on the room's communication floor, and the amount of power directed, must not cause the lighting level to be in violation of the acceptable limits. Although generating a small size beam leads to high received optical power, this reduces the CGHs coverage area, which restricts user mobility. In contrast, a large size beam increases the CGHs coverage area; but, this offers low received optical power. Hence, a beam with an optimal size must be generated for the required coverage and mobility levels. Because of the light sources distributed on the room's ceiling, the room's communication floor was split into eight 2 m × 2 m surface areas (see Fig. 2). In addition, the generated beam size is chosen to be 2 m × 2 m. Choosing the closest light unit to the optical receiver leads to coverage of all possible locations in the room' communication floor.

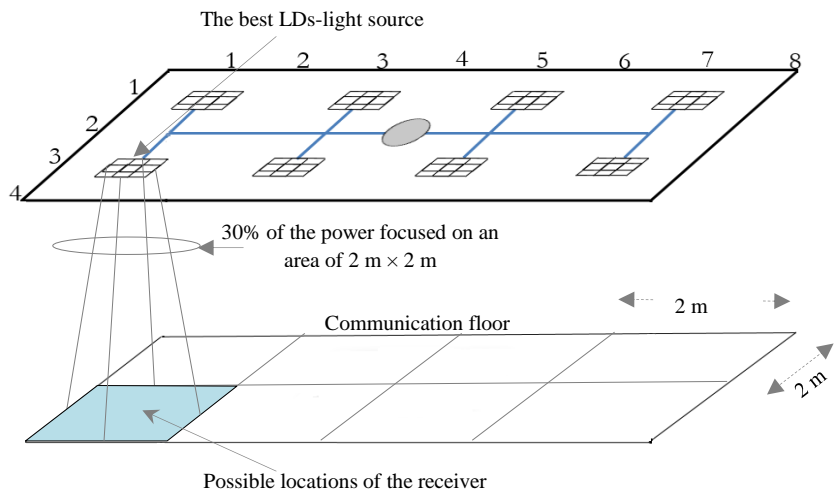

Fig. 2: Size of the beam generated by the CGHs.

To determine the amount of optical power that should be generated by the CGHs from the best light source without violating the acceptable lighting levels in the room, we examined different amounts of optical power that may be directed from the best light source by the CGHs such as 20%, 30% and 40%. These values were examined while taking into account that the best light source may be the one positioned at one of the room's corners (we considered the light source that was located at 1 m, 1 m, 1 m as the best light source) since the minimum lighting happens at the room's corners. The results revealed that up to 30% of the total optical power from the best light source can be directed to the 2 m × 2 m surface area while keeping the lighting at the level established by the standard (i.e. 300 lx [25]). Fig. 3 illustrates the lighting level distribution on the room's communication floor. As seen, before generating the beam, the lowest value of lighting was 338 lx, which met the standards requirements (i.e. more than 300 lx). The lowest illumination levels while using 20%, 30% and 40% of the total optical power of the best light source at the corner of the room were 314 lx, 302 lx and 286 lx, respectively. So, we selected 30% of the total optical power from the light source to keep the level of the lighting at an appropriate value.

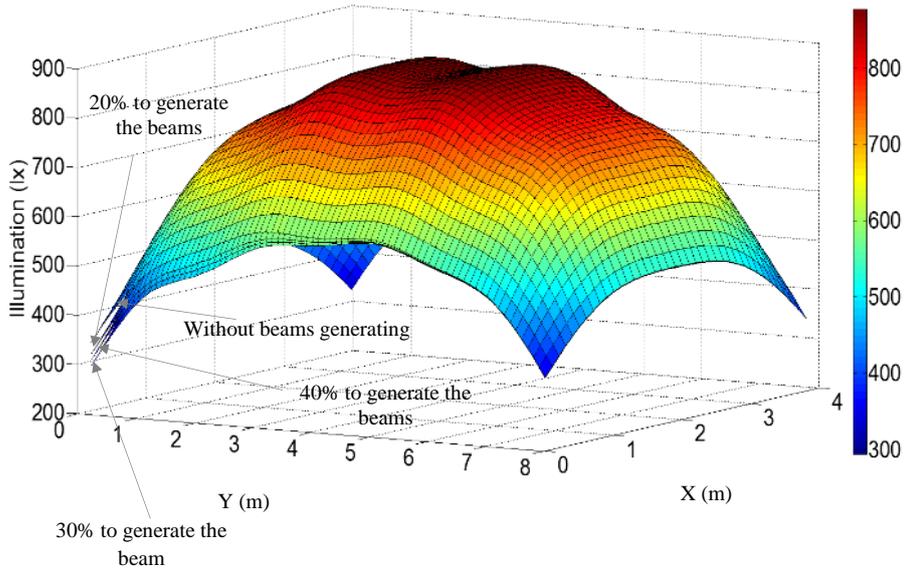

Fig. 3: The distribution of illumination on the communication floor.

## 4 Results

Our proposed VLC system's performance is examined in two rooms: room A (the idealistic room) and room B (the realistic room). We utilized a simulation tool similar to that in [32], [33]. In this simulation tool, the impulse responses were determined, and consequently, the optical path loss (*PL*), delay spread (D) and 3dB bandwidth of the channel were evaluated. The VLC systems were examined while the optical receiver moved along the *x*-axis and *y*-axis in steps of 1 m. Because of the symmetry of room A, we obtained the results on the *y*-axis and along *x* = 1 m and *x* = 2 m whereas the results were calculated along the *y*-axis and on *x* = 1 m, *x* = 2 m and *x* = 3 m for room B. For comparison purposes, we report the results of our proposed system without and with CGHs, and for both systems, one transmitter was used to send the data.

### 4.1 Results in the Empty Room (room A)

The delay spread related with the impulse response is given as [34]:

$$D = \sqrt{\frac{\sum (t_i - \mu)^2 P_{ri}^2}{\sum P_{ri}^2}} \quad (5)$$

here $t_i$ denotes the delay time related with the received optical power $P_{ri}$ and $\mu$ denotes the mean delay written as:

$$\mu = \frac{\sum t_i P_{ri}^2}{\sum P_{ri}^2} \quad (6)$$

The delay spread of the indoor VLC system without using the CGH and using the CGH is shown in Fig. 4. The results were calculated while the optical receiver moved at *x* = 1 m and *x* = 2 m along the *y*-axis in room A. The results show that using the CGH for the indoor VLC system leads to reduced delay spread and consequently reduced ISI at higher data rates. As shown in Fig. 4, the maximum delay spread value was 0.27 ns for the VLC system without the CGH whereas it was 0.09 ns for the VLC system that use the CGH. The improvement is due to the 30% of the total optical power of the best transmitter directed to the 2 m × 2 m surface area. This increases the received optical power associated with the LOS components and decreases the effect of diffuse reflections from the surface reflectors of the room.

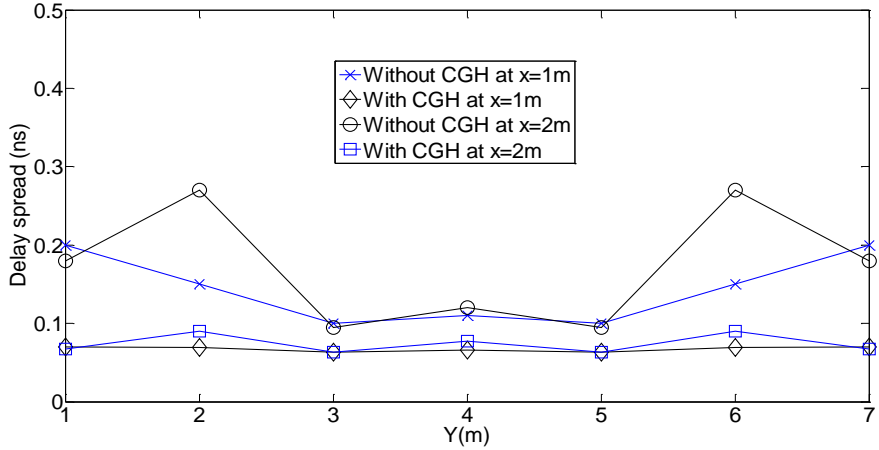

Fig. 4: Delay spread of the mobile user without and with the CGH in room A along $x = 1$ m and $x = 2$ m.

Table I shows the 3dB bandwidth of the indoor VLC channel without the CGHs and with the CGHs while the mobile user moved along the $y$-axis at $x = 1$ m and $x = 2$ m in room A. As can be seen, using the CGHs in the VLC system leads to increase in the 3dB bandwidth compared to the VLC system without the CGHs. The minimum 3 dB bandwidth of the VLC channel without the CGHs was 1.21 GHz. In contrast, the lowest 3dB channel bandwidth of the VLC system with the CGHs was 2.2 GHz. Thus, with a bandwidth equal to 0.7 times the data rate (for OOK modulation) [35], the highest data rate that is supported by the VLC system without and with the CGHs is 1.73 Gb/s and 3.15 Gb/s, respectively.

TABLE I
3dB CHANNEL BANDWIDTH WITHOUT AND WITH CGH ALONG THE Y-AXIS AND AT X = 1 m AND X = 2 m.

| 3 dB channel bandwidth (GHz) | | | | | |
|---|---|---|---|---|---|
| **Receiver locations (m)** | **without CGH** | **with CGH** | **Receiver locations (m)** | **without CGH** | **with CGH** |
| 1, 1, 1 | 1.25 | 2.86 | 2, 1, 1 | 1.38 | 2.98 |
| 1, 2, 1 | 1.75 | 2.89 | 2, 2, 1 | 1.21 | 2.2 |
| 1, 3, 1 | 2.1 | 3.1 | 2, 3, 1 | 2.3 | 3.17 |
| 1, 4, 1 | 1.9 | 3.03 | 2, 4, 1 | 1.67 | 2.6 |
| 1, 5, 1 | 2.1 | 3.1 | 2, 5, 1 | 2.3 | 3.17 |
| 1, 6, 1 | 1.75 | 2.89 | 2, 6, 1 | 1.21 | 2.2 |
| 1, 7, 1 | 1.25 | 2.86 | 2, 7, 1 | 1.38 | 2.98 |

### 4.2 Effect of Shadowing and Signal Blockage on the Proposed System

To study the influence of obstacles on our proposed VLC system, the analysis was carried out in room B. In room B, we took into account the influence of signal blockage attributed to mini cubicles, door, windows, furniture, multipath propagation and user mobility (see Fig. 1 (b)). Because of the asymmetry of room B, results were calculated while the mobile user moved along the $y$-axis on the lines $x = 1$ m, $x = 2$ m and $x = 3$ m.

Fig. 5 shows the received optical power gain due to the use of CGHs when the mobile receiver moved along the $y$-axis and at $x = 1$ m, $x = 2$ m, and $x = 3$ m in steps of 1 m in room B. This gain is expressed as:

$$Power\ gain\ (dB) = Pr_{CGH}(dB) - Pr(dB) \tag{7}$$

where $Pr_{CGH}(dB)$ is the received optical power using CGH and $Pr(dB)$ is the received optical power without the CGH. It can be seen that due to the use of the CGH, the minimum gain in optical received power was equal to 3 dB, 2.2 dB and 2.8 dB along the $x = 1$ m, $x = 2$ m and $x = 3$ m, respectively. Thus, by using the CGH, the quality of the connection improves without increasing the transmitted power.

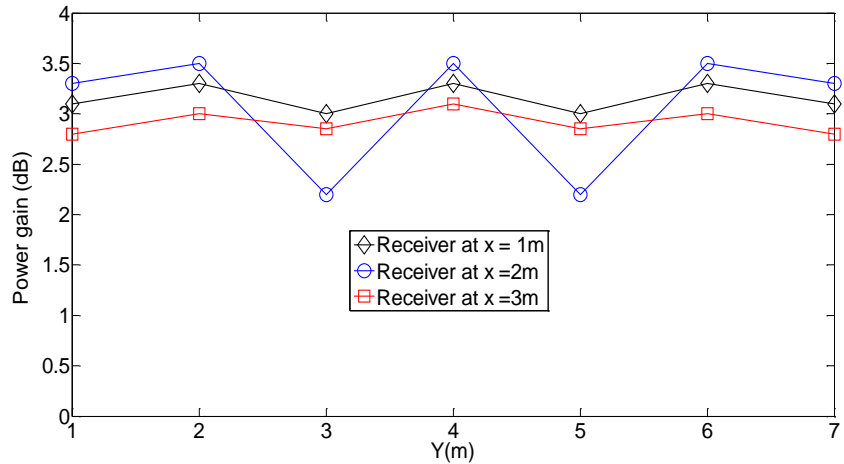

Fig. 5: Received optical power gain when the mobile user was walking along the y-axis and at x = 1 m, x = 2 m and x = 3 m in room B.

## 5 Conclusions

An indoor VLC system that utilizes CGHs was proposed in this work. This VLC system was designed to overcome one of the main impairments in indoor VLC systems, namely ISI caused by multipath dispersion. The CGH was utilized to direct a single fixed beam from the best light source to a surface area of 2 m × 2 m, which improved the 3-dB bandwidth of the indoor VLC channel and increased the received optical power. We considered the acceptable lighting levels and the impact of our approach on lighting levels. Thus, we optimized the amount of optical power that depicted by the CGHs, which was 30% of the total power of the best light source in our room setup. In addition, we proposed a "select the best light source" algorithm in which the best light source was assigned to the optical receiver without needing to know the receiver location, which simplifies the design of the VLC system. We evaluated the performance of our proposed system in two different room settings in the presence of multipath propagations and mobility. Our proposed system decreased the delay spread of the traditional VLC systems (VLC systems that do not use CGHs) and improved the 3dB channel bandwidth, which helps support higher data rates.